\documentclass[fleqn,usenatbib,letters]{mnras}

\usepackage{newtxtext,newtxmath}

\usepackage[T1]{fontenc}

\DeclareRobustCommand{\VAN}[3]{#2}
\let\VANthebibliography\thebibliography
\def\thebibliography{\DeclareRobustCommand{\VAN}[3]{##3}\VANthebibliography}

\usepackage{graphicx}	
\usepackage{amsmath}	
\usepackage{amssymb}	
\usepackage{xcolor}
\usepackage{xspace}

\newcommand{\gaia}{{\it Gaia}\xspace}
\newcommand{\HST}{{\it Hubble Space Telescope}\xspace}
\newcommand{\VLT}{{\it Very Large Telescope}\xspace}

\defcitealias{Kluter2018b}{K18b}
\defcitealias{Kluter2019}{K19}

\title[spurious microlensing events]{Predictions of Gaia's prize microlensing events are flawed}

\author[P. McGill et al.]{
Peter McGill$^{1}$\thanks{Email: pm625@cam.ac.uk (PM)}, 
Andrew Everall$^{1}$,
Douglas Boubert$^{2}$ and
Leigh C. Smith$^{1}$
\\
$^{1}$Institute of Astronomy, University of Cambridge, Madingley Rd, Cambridge CB3 0HA, UK\\
$^{2}$Magdalen College, University of Oxford, High Street, Oxford OX1 4AU, UK
}

\date{}

\pubyear{2020}

\begin{document}

\label{firstpage}
\pagerange{\pageref{firstpage}--\pageref{lastpage}}
\maketitle

\begin{abstract}
Precision astrometry from the second \gaia data release has allowed astronomers to predict 5,787 microlensing events, with 528 of these having maximums within the extended \gaia mission (J2014.5 - J2026.5). Future analysis of the \gaia time-series astrometry of these events will, in some cases, lead to precise gravitational mass measurements of the lens. We find that 61\% of events predicted during the extended Gaia mission with sources brighter than G = 18 are likely to be spurious, with the background source in these cases commonly being either a duplicate detection or a binary companion of the lens. We present quality cuts to identify these spurious events and a revised list of microlensing event candidates. Our findings imply that half of the predictable astrometric microlensing events during the \gaia mission have yet to be identified. 
\end{abstract}




\section{Introduction}

The astrometric signatures of microlensing events offer singular opportunities for direct gravitational probes of the fundamental properties of stars and stellar remnants, whether that be the mass of single objects \citep[e.g][]{Paczynski1995,Miralda-Escude1996,Rybicki2018,Kains2017} or characteristics of their populations \citep[e.g][]{Dominik2000, Belokurov2002,Lam2020}. Microlensing occurs when a massive object (the lens) aligns closely with a distant background source as seen by an observer. Two images of the source are formed, resulting in an apparent brightening of the source \citep[photometric microlensing  --][]{Paczynski1986}, and apparent deflection of its position \citep[astrometric microlensing -- ][]{Hog1995,Miyamoto1995,Walker1995}. 

While photometric signatures of microlensing events are routinely detected by large scale monitoring surveys of the Galactic bulge (e.g the Optical Gravitational Lensing Experiment, OGLE - \citealt{OGLEIV2019}, or the Korea Microlensing Telescope Network, KMTNet - \citealt{KMTnet2016}), the detection of astrometric effects is still rare. This is due to the lack of large scale astrometric monitoring surveys which publish time-series astrometry.  Despite this, astrometric microlensing events can still be found. It is possible to predict lens-source alignments ahead of time \citep{Refdal1964}, if the positions, proper motions, and parallaxes of both the background source and lens are known. The strength of the microlensing signal can be computed based on a mass estimate for the lens, allowing targeted follow-up campaigns to be organized to observe the event.

As astrometric catalogues increased in quality and number, many searches for predicted events were carried out \citep{Feibelman1966,Salim2000,Proft2011,Lepine2012,Sahu2014,Harding2018, McGill2018}. To date, only two events from these predictions have been detected. \cite{Sahu2017} used the \HST to measure the mass of white dwarf Stein 2051 b to 8\% precision. \cite{Zurlo2018} used the \VLT to obtain a 40\% precision mass measurement of Proxima Centauri. The advent of astrometric data at an unprecedented precision and volume from the second \gaia data release \citep[GDR2 -][]{GDR22018} reignited interest in predicting microlensing events. Searches by many studies resulted in precise predictions of 5,787 microlensing events occurring over the next century \citep{Bramich2018,Mustill2018,Kluter2018a,Kluter2018b,Bramich2018b,McGill2019a}. In addition to predicting future microlensing events, events occuring over \gaia's observation baseline were presented with the view that they could be analysed when \gaia releases time-series astrometry \citep[e.g.][]{Kluter2019}.

Can we trust these predictions? How many of these events will \gaia observe? In this letter we examine these questions in detail.

\section{The predicted events}

We analyse the predicted microlensing events found by searches solely using GDR2 \citep{Bramich2018,Mustill2018,Kluter2018a,Kluter2018b,Bramich2018b,McGill2019a}, giving us a total sample of 5,787 distinct events caused by 4,436 lenses. Although many of these studies predict some of the same events, there are key differences. \cite{Mustill2018} searched for photometric events caused by lenses within $100\;\mathrm{pc}$ over the next 20 years. \cite{Kluter2018a} presented two on-going astrometric events which at the time required immediate follow up. \cite{Bramich2018} presented a catalogue of photometric and astrometric events with maximums during the extended \gaia mission (J2014.5 - J2026.5). \cite{Kluter2018b} (hereafter \citetalias{Kluter2018b}), presented a catalogue of predicted photometric and astrometric events with maximums between J2014.5 and J2065. \cite{Bramich2018b} presented a catalogue of photometric and astrometric events with maximums between the end of the extended \gaia mission and the end of the Century (J2026.5 - J2100.0). Finally, \cite{McGill2019a} presented two photometric events which required immediate follow up in J2019. Each of these studies used different event detectability criteria, lens and source selection criteria, and lens mass estimates, which resulted in different sets of events. In most cases where two different studies found the same event the predictions were consistent. Searches using GDR2 in combination with other catalogues have also occurred \citep{Ofek2018,Nielsen2018,McGill2019b}, but we do not consider them further in this work.

\section{The intriguing case of G123-61}

At a first glance, the predicted microlensing event by the lens G-123-61A (lens - \gaia DR2 1543076475509704192, $G=12.9$; source - \gaia DR2 1543076471216523008, $G=13.1$) looks like a promising candidate. This event, predicted by \citetalias{Kluter2018b}, peaked on J2016.311 with a predicted maximum astrometric deflection of $\sim$0.45 mas. \cite{Kluter2019} conclude that with time series astrometry from \gaia this event should permit a mass determination of G-123-61A to 24\% precision. The quality of this estimated constraint is largely due to the high apparent brightness of the source. Crucially, this would allow \gaia to obtain high precision single epoch measurements of the astrometric deflection \citep{Rybicki2018}. 

\begin{figure}
    \centering
    \includegraphics[width=1.\linewidth]{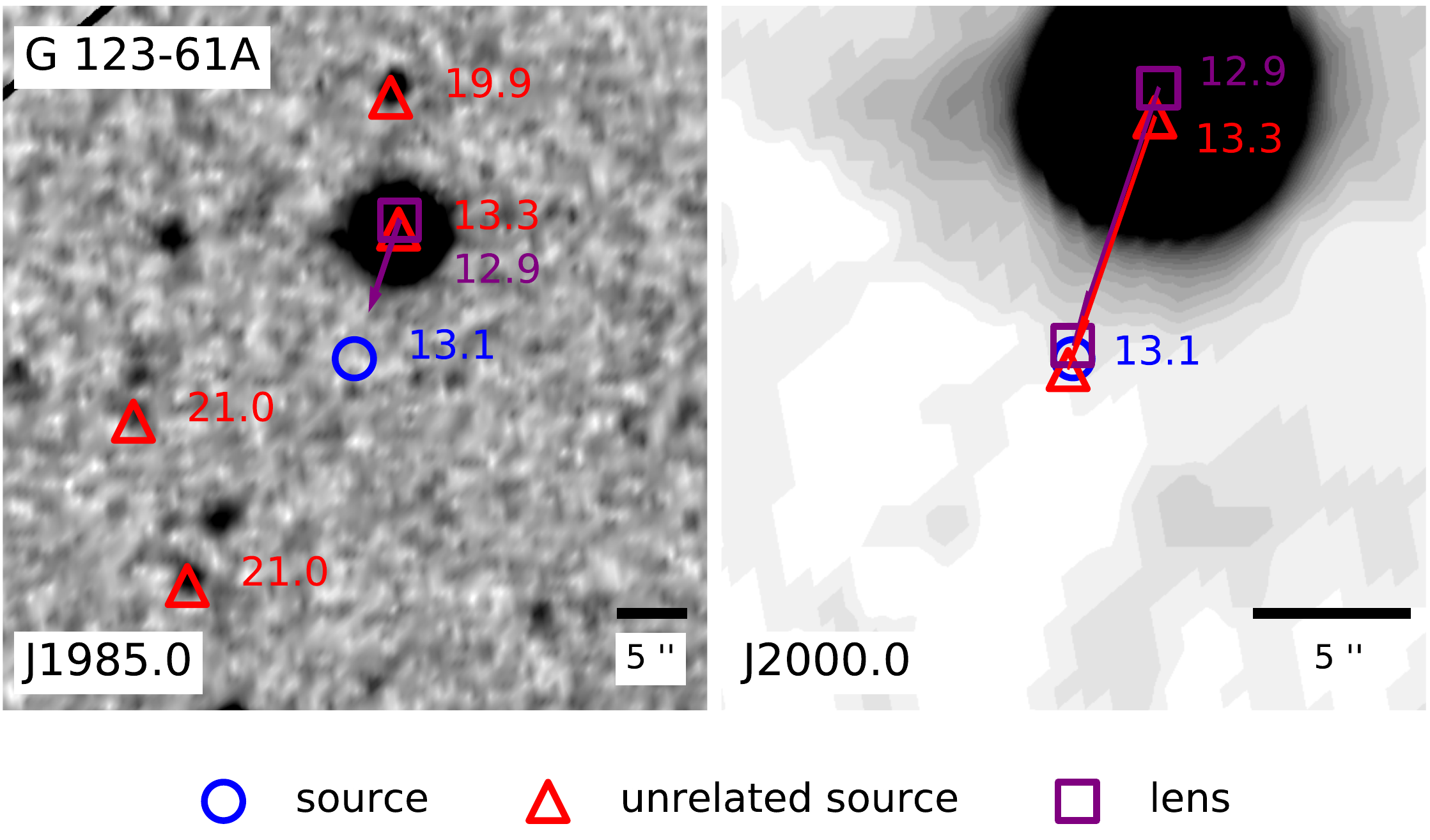}
    \caption{Images of the region surrounding the predicted microlensing event caused by the lens G123-61A showing that the source is not present. Arrow shows the direction of the proper motion of the lens. Annotated text gives the GDR2 $G$ magnitudes. North is in the upwards vertical direction in both images. \textbf{Left:} DSS-blue image (epoch $\sim$ J1985.0) with positions projected to the image epoch if the source has a GDR2 proper motion, otherwise the position is at the GDR2 reference epoch of J2015.5. 
    \textbf{Right:} 2MASS $K_s$-band image (epoch $\sim$J2000.0) of the event with positions shown at both the GDR2 reference epoch of J2015.5 and the 2MASS image epoch.}
    \label{fig:cutouts}
\end{figure}

Fig. \ref{fig:cutouts} (left) shows a DSS2 cutout image of the field around the G-123-61A event. The positions are the projections of the GDR2 astrometric solution back to the epoch of the DSS2 image ($\sim$J1985). The source in the event does not have a GDR2 proper motion and therefore cannot be projected back to the DSS2 epoch -- its position is at the GDR2 epoch of J2015.5. It is clear from Fig. \ref{fig:cutouts} that a thirteenth magnitude source is not present in the image at the GDR2 J2015.5 position which was used in the event's prediction, nor is there an unaccounted thirteenth magnitude source elsewhere in the image.

\citetalias{Kluter2018b} predict a second microlensing event with \gaia DR2 1543076471216523008 as the source by the lens G-123-61B (\gaia DR2 1543076475514008064, $G=13.3$). This event peaked on J2014.76 with a predicted maximum astrometric deflection of $\sim$1 mas. \citet{Kluter2019} predict this will enable \gaia to measure G-123-61B's mass to 37\% precision. The lens stars for these two predicted events (G-123-61A and G-123-61B) have similar GDR2 positions, proper motions and parallaxes, and have consequently been assigned to a binary system by SIMBAD. 
Whilst there is no reference in the literature to G-123-61 being a binary prior to GDR2, there is some weak indication of it. The radial velocity from \citet{GDR22018} of G-123-61A is $-28.69 \pm 2.80\;\mathrm{km}\;\mathrm{s}^{-1}$ based on eight measurements, whereas \citet{Reid1995} measures the radial velocity to be $3.5\pm10\;\mathrm{km}\;\mathrm{s}^{-1}$. These values are mildly inconsistent and provide weak evidence that G-123-61 is indeed a binary, and suggests that the two GDR2 lens detections could be genuine. Regardless of this, Fig. \ref{fig:cutouts} shows that for both of these events the source star is suspect. G-123-61B can be seen in Fig. \ref{fig:cutouts} as the 13.3 magnitude GDR2 source marked as unrelated to the G-123-61A event.

\begin{figure*}
	\centering
    \includegraphics[width=1.\linewidth,trim=0 0 0 0, clip]{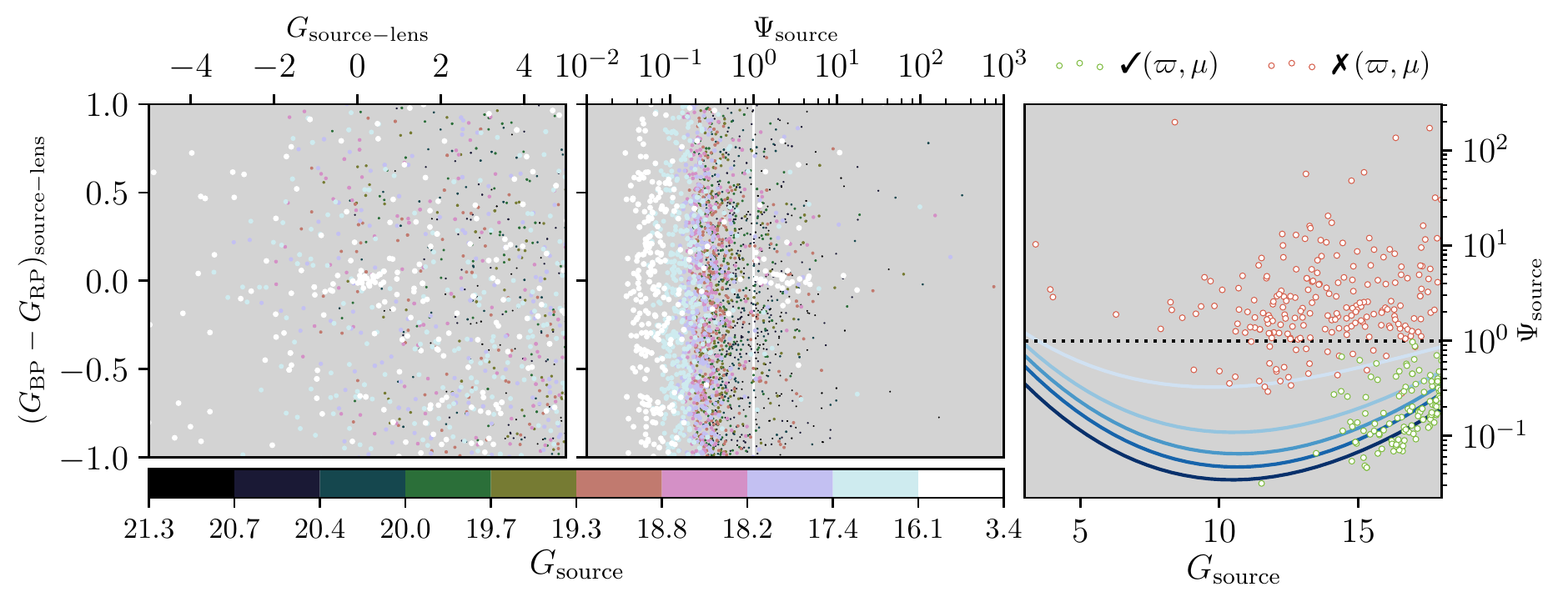}
	\caption{Many of the predicted microlensing events have sources with suspect properties. \textbf{Left:} The magnitude and colour difference between the source and lens for the microlensing events predicted with GDR2 astrometry, with the size and colour of the marker indicating the source magnitude. There is a cluster of photometrically identical lens-source pairs. \textbf{Middle:} Same schema as the left panel but showing the astrometric quality indicator $\Psi$ for the source, as defined by Eq. \ref{eq:psi}, with $\Psi\leq1$ being required for the source to have a parallax and proper motion in GDR2. The sources in the photometric cluster are all missing these quantities. \textbf{Right:} For events peaking prior to J2026.5, we show the magnitude of the source versus $\Psi$, with those sources having/missing GDR2 parallaxes and proper motions circled in green/red. The blue lines gives the 50\%, 80\%, 90\%, 95\% and 98\% percentiles of $\Psi$ for all stars in GDR2 binned by $G$ magnitude (smoothly interpolated for illustrative purposes). Most of the sources in predicted microlensing events peaking before J2026.5 lie above the 98\% line.}
	\label{fig:diagnostic}
\end{figure*}

A troubling aspect of these two events is that at J2015.5 the source lies in between the two suggested lenses G-123-61A and B. Furthermore, both the lenses and the source are similar magnitudes, and the source ($G_{\mathrm{BP}}-G_{\mathrm{RP}}=2.67$) has an almost identical colour to the lens G-123-61A ($G_{\mathrm{BP}}-G_{\mathrm{RP}}=2.66$, the other lens does not have GDR2 colour photometry). The rest of this paragraph is an abridged explanation provided by the \gaia Helpdesk in response to our queries. Using the scanning law from \citet{Boubert2020} and the astrometric gaps during GDR2 provided by DPAC (\url{https://www.cosmos.esa.int/web/gaia/dr2-data-gaps}), we were able to predict that this system should have been observed 65 times during the 22 months of GDR2. Each of these observations should have resulted in a detection of each component of the thirteenth magnitude binary \citep{Boubert2020b}, giving a total of 130 detections. It is not a coincidence that the total detections (\textsc{matched\_observations}) of G-123-61A (54), G-123-61B (42) and the source (34) also sums to 130. We deduce that when the \gaia DPAC clustering algorithm was merging detections into stars, a fraction of the 65 detections of the two components of the binary were mistakenly combined into a spurious third star. It is likely that the proper motion of the system being in the same direction as the separation of the two components made this a difficult system for the clustering algorithm. The donated astrometric detections from the two components of the binary will lead to a problematic astrometric solution for the spurious source, thus explaining why the source does not have an astrometric solution despite being a thirteenth magnitude star. We can thus conclude that the source is spurious and that these two predicted microlensing events did not occur.


While we expect cases like that of G-123-61 to be rare, the trouble does not end there. Inspection of imaging around several other events revealed further missing sources, with one example being the event  caused by the lens LP 701-45 (lens - \gaia DR2 2610954226042154624, $G=14.8$; source - \gaia DR2 2610954226041533696, $G=17.3$) predicted by \citetalias{Kluter2018b} to peak on J2022.36 with a deflection $\sim$0.1 mas. The \gaia uncertainties on the right ascension and declination of the source are highly degenerate with a correlation of 99.8\%, suggesting that the 2D astrometric pipeline was attempting to fit points lying along a line. We suspect that the source in this event is truly a binary companion of the lens, explaining why it does not appear in legacy DSS2 imaging. In many cases we were not able to tell by eye from the DSS2 imaging whether the source was visible at the J2015.5 position because the lens was not well enough separated at $\sim$J1985, but our suspicions were sufficiently raised that we decided to investigate further.

\section{Diagnostic test for spurious microlensing candidates}

Typical predictable astrometric microlensing events are caused by nearby -- and therefore bright and high proper motion -- lenses and more distant -- and therefore usually fainter -- background sources. This is because small observer-lens distances cause larger astrometric signals \citep{Dominik2000}, and high proper motion objects are more likely to align with a source over a given time. With this in hand our n\"aive expectation of a source-lens pair is that the lens should be a bright star and that the source should be a randomly picked star in the background (and thus resemble a `typical' star in GDR2). If the properties of the source are unusual or are similar to that of the lens, then we should question whether the predicted microlensing source-lens pair is real. 

In Fig. \ref{fig:diagnostic} (left) we show the difference in colour between the source and lens versus the difference in their magnitudes. Surprisingly, there is a cluster of lens-source pairs with bright sources where the source colour is within $0.1\;\mathrm{mag}$ of the lens colour. Given that the most observable microlensing events are those with bright sources, we decided to investigate the astrometric properties of these sources to see if they are consistent with being genuine stars.

Stars only have reported parallaxes and proper motions in GDR2 if they satisfy the three criteria given by \citet{Lindegren2018}: $G\leq21$, $\textsc{visibility\_periods\_used} \geq 6$, and $\textsc{astrometric\_sigma5d\_max} \leq(1.2\;\mathrm{mas})\times \gamma(G)$.
%
The function $\gamma(G) = \textrm{max}[1, 10^{0.2(G-18)}]$ is flat for $G\leq18$ and then transitions to exponential growth. The quantity \textsc{visibility\_periods\_used} is the number of time resolved clusters of detections used in the astrometric pipeline and is required to be at least six to ensure a long enough baseline for the astrometric solution. The quantity \textsc{astrometric\_sigma5d\_max} is the square root of the largest singular value of the scaled $5\times5$ covariance matrix of the astrometric parameters, and so is comparable to the semi-major axis of a position error ellipse \citep{Lindegren2018}. The third cut can be interpreted as requiring that the astrometric uncertainty should not be unusually large for a star of that magnitude. We define the quantity
\begin{equation}
    \Psi = \frac{\textsc{astrometric\_sigma5d\_max}}{(1.2\;\mathrm{mas})\times \gamma(G)}, \label{eq:psi}
\end{equation}
such that if $\Psi > 1$ then the source will fail the third cut. We show $\Psi$ versus the source-lens colour difference in the middle panel of Fig. \ref{fig:diagnostic}. The cluster of source-lens pairs with identical photometry in the left panel stands out in the middle panel, with all of these sources having $\Psi>1$. This is highly unusual for sources brighter than $G=16.1$ -- in Fig. \ref{fig:diagnostic} most of these sources have $\Psi<0.1$ -- and so we can conclude that these sources are atypical and thus concerning. All of the events in the cluster were identified by \citetalias{Kluter2018b} and were predicted to have their peak prior to J2026.5. We emphasise that the left and middle panels of Fig. \ref{fig:diagnostic} only show the sources with colour photometry. There is a few-fold larger group of sources without colour photometry that have $G_{\mathrm{source}}<G_{\mathrm{lens}}$, $\Psi\gtrsim 0.3$ and a peak prior to J2026.5, and thus the cluster in Fig. \ref{fig:diagnostic} is only a subset of the phenomenon. We conjecture that the reason only those sources in the extended cluster with $G_{\mathrm{source}}\approx G_{\mathrm{lens}}$ have colour photometry is the colour excess cut applied by DPAC, $E=(F_{\mathrm{BP}}+F_{\mathrm{RP}})/F_G<5$. The colour photometry measured for these sources is dominated by the flux from the nearby lens and so $F_{\mathrm{BP/RP,source}}\approx F_{\mathrm{BP/RP,lens}}$, and thus $E_{\mathrm{source}}\gg5$ unless the source and lens have similar magnitudes.

In Fig. \ref{fig:diagnostic} (right) we show the sources with $G<18$ and a predicted astrometric microlensing signal peak before J2026.5. Most of these sources are astrometrically-unusual compared to typical GDR2 sources, having an \textsc{astrometric\_sigma5d\_max} in the top 2\% of stars at that magnitude. Those sources with extreme astrometric errors are mostly those without a 5D astrometric solution, while those with 5D astrometry are representative of the DR2 source catalogue. We propose that events with sources brighter than G=18 should only be considered reliable if the source has a published 5D astrometric solution.



\begin{figure}
	\centering
    \includegraphics[width=1.\linewidth,trim=0 0 0 0, clip]{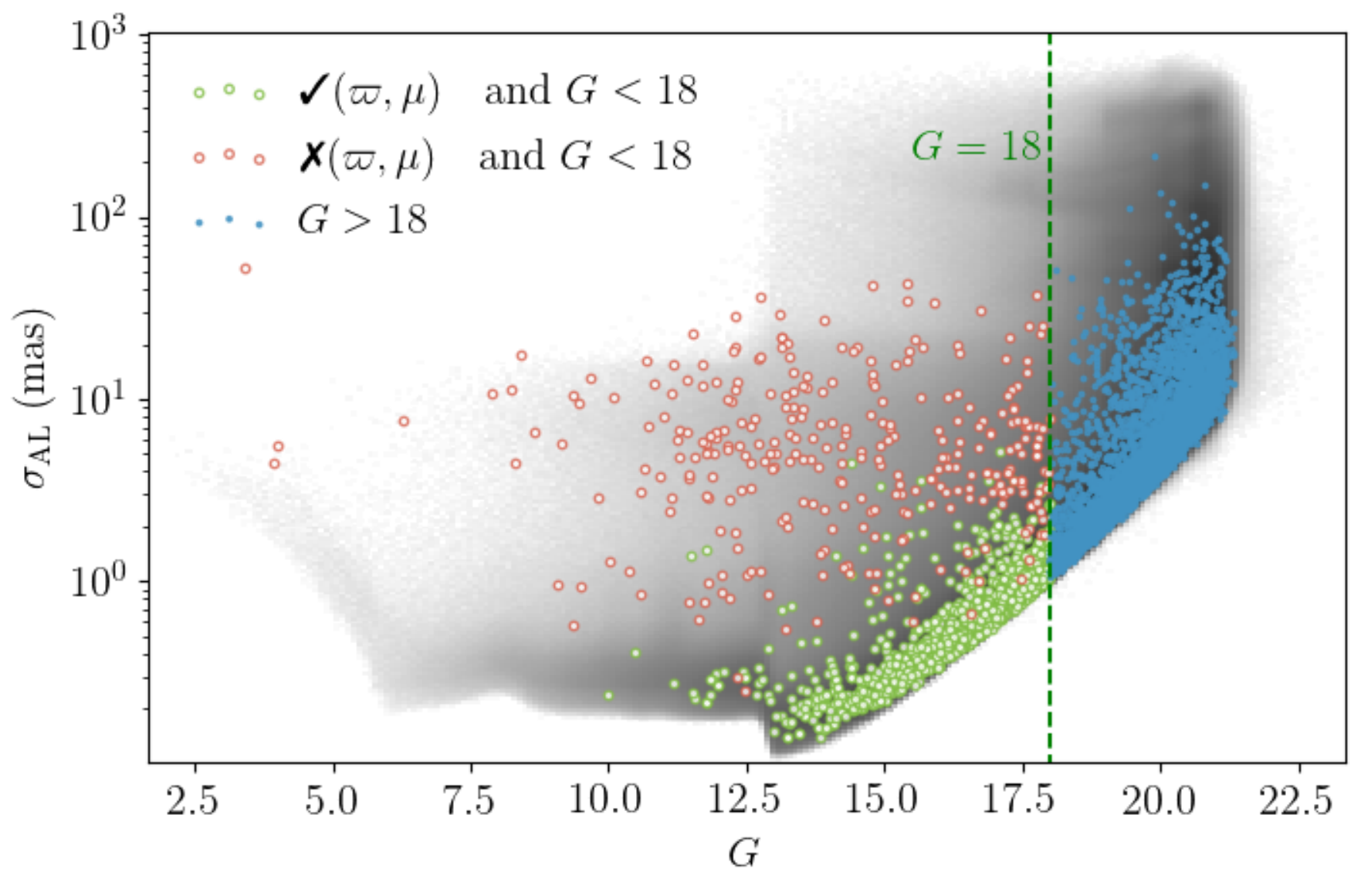}
	\caption{The well-behaved microlensing candidate sources brighter than $G=18$ (green points) have along-scan (AL) astrometric errors per observation that are typical of the GDR2 population (grey-scale background, log-normalised) whilst spurious source candidates (red points) have systematically higher AL error. At dimmer magnitudes, the expected position error of well-behaved sources is high enough that the all event sources (blue points) trace the bulk distribution.}
	\label{fig:sigmaal}
\end{figure}

Two factors determine whether a bright source will have a published 5D astrometric solution. Firstly, fewer detections will reduce the likelihood of having the requisite six visibility periods as well as making it more likely that $\Psi>1$, because fewer measurements constraining the astrometric solution will increase the astrometric uncertainty. For each source with $G\leq18$ and a microlensing peak prior to J2026.5, we calculated the ratio of astrometric detections $k$ (\textsc{astrometric\_matched\_observations}) to the predicted number of astrometric observations $n$ computed using the scanning law of \citet{Boubert2020} and the published gaps from DPAC. The $1\sigma$ interval $k/n\in(22.9,60.0)\%$ for the sources without 5D astrometry is significantly lower than the $1\sigma$ interval $k/n\in(65.0, 95.8)\%$ for the sources with 5D astrometry. Secondly, an increase in either the astrometric uncertainty of the centroiding of individual detections or the scatter between the centroids around the single source astrometric fit will increase the reported astrometric error. We estimated the typical centroiding uncertainty from the harmonic mean of the equatorial positional uncertainties ($\sigma_\mathrm{AL}=\sqrt{\nu/(1/\sigma_\alpha^2+1/\sigma_\delta^2)}$ where $\nu$ is \textsc{astrometric\_n\_good\_obs\_al}), and show in Fig.~\ref{fig:sigmaal} that the sources without a 5D astrometric solution have enhanced centroid error compared to both the predicted microlensing sources with 5D astrometry and to the bulk of \gaia DR2. We note some of the sources at magnitudes dimmer than $G=18$ are likely to be spurious events, however these are difficult to distinguish from the main population as astrometric error increases with magnitude.


\begin{figure}
	\centering
    \includegraphics[width=1.\linewidth,trim=0 0 0 0, clip]{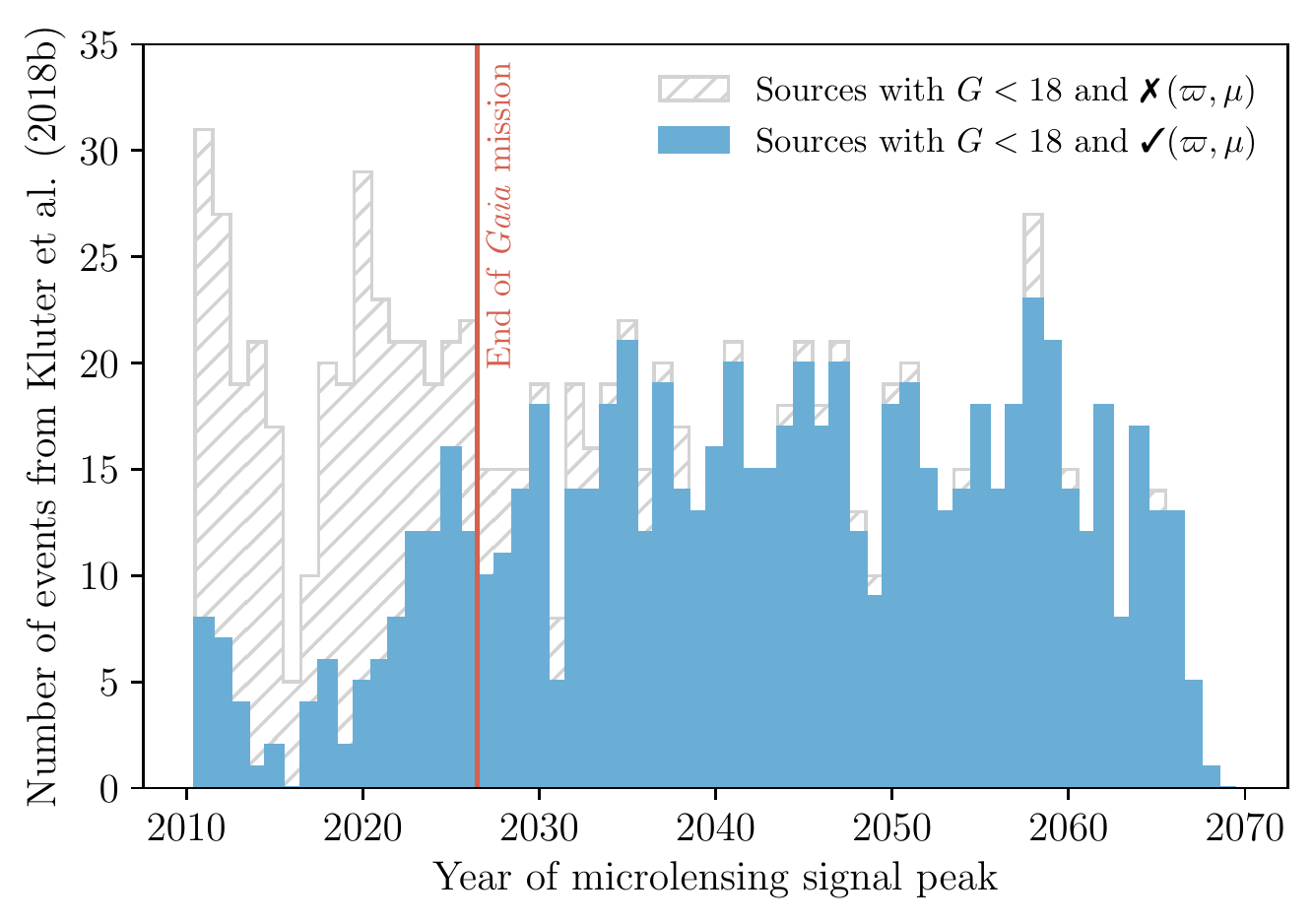}
	\caption{The number of microlensing events predicted per year by \citetalias{Kluter2018b}. If we do not apply any additional quality cuts then the rate peaks near present day. Applying our astrometric quality cut $\Psi\leq1$ removes most of the predicted events during the horizon of the extended \gaia mission.}
	\label{fig:timeline}
\end{figure}


Requiring 5D astrometry resolves another troubling property of the current sample of predicted microlensing events: the microlensing rate appears to peak at present day. In Fig. \ref{fig:timeline} we show the number of microlensing events with $G_{\mathrm{source}}<18$ predicted per year by \citetalias{Kluter2018b}. We restrict ourselves to \citetalias{Kluter2018b} because it is easier to interpret the rate if we only need to consider one set of selection cuts. Our expectation is that there should be a fairly constant rate of microlensing events, but, as noted by \citetalias{Kluter2018b}, there should be a deficit of events around the GDR2 epoch of J2015.5 because the lens and source are less likely to be resolved by \gaia at their point of closest approach. Paradoxically, the rate of predicted microlensing events peaks either side of a dip at J2015.5. This is strong evidence that these events are unlikely to be real. If we remove sources without 5D astrometry then the microlensing event rate matches our expectations, with that cut predominantly removing sources with peaks near to J2015.5. We use the number of events per year between J2034.5 and J2055.5 to infer that the Poisson rate of \citetalias{Kluter2018b}-type astrometric microlensing events is $16.0_{-0.8}^{+0.9}\;\mathrm{events}\;\mathrm{yr}^{-1}$, assuming the inverse-square-root Jeffrey's prior on the rate. Marginalising over this rate,  we find that there should be $191_{-17}^{+18}$ events during the twelve year horizon of the extended \gaia mission, only 85 of which found by \citetalias{Kluter2018b} meet our criteria for reliable identification. Over half of the astrometric microlensing events that will be detected by \gaia are yet to be identified.
      

 Not all of the events during the \gaia mission predicted by \citetalias{Kluter2018b} will result in useful mass measurements of the lens. Filtering events by our new astrometric cut changes the outlook for astrometric lensing events with \gaia. Of the 513 events predicted by \citetalias{Kluter2018b} to peak between J2014.5 and J2026.5, only 260 have a source with 5D astrometry. Bright sources suffer an even higher attrition rate with only 85 of the 227 events with $G_{\mathrm{source}}<18$ surviving. This will impact the prospects of high-precision lens mass measurements with \gaia. \citet{Kluter2019} identified that 62 of their predicted single lens-source events could lead to mass measurements of the lens with precision better than 100\%, but we find that only 28 of these are likely to occur. All seven events with an estimated precision on the mass of the lens below 10\% with the full ten year \gaia data are unlikely to occur. Notably the 5.4\% mass estimate of 75 Cancri is unlikely to be realised. We conclude that many of these spurious events are likely caused by the source star being a detection of a binary companion of the lens.



The reliability of \citetalias{Kluter2018b}'s predictions increases significantly beyond J2026.5 . Requiring 5D astrometry for sources with G < 18 only eliminates 46 of the total 3221 events. All of the events predicted by \cite{Kluter2018a},  \cite{Mustill2018} and \cite{McGill2019a} survive. Only 2 of the 76 events predicted to happen over the extended \gaia mission by \cite{Bramich2018} are eliminated. Finally, 37 of the 2509 events predicted to happen between J2026.5-J2100.0 by \cite{Bramich2018b} are eliminated by our cuts. The bulk of the likely spurious events occur around the GDR2 reference epoch (J2015.5) and therefore make up a large fraction of the events predicted to happen during the extended \gaia mission.

Inspection of the DSS2 imaging data of the two events predicted by \cite{Bramich2018} but eliminated by our cuts reveal that the source and lens are clearly present and thus that these events are likely real. To ensure that this is a rare occurrence and our proposed quality cuts don't eliminate many real events, three of the authors independently eyeballed the 231 events with $G<18$ and predicted peak between J2014.5 and J2026.5. We used the ESAsky (\url{https://sky.esa.int/}) online observation tool with images from DSS2, 2MASS, SDSS9 and AllWISE to determine whether an event is plausible. Of the 89 events for which the source has a 5D astrometric solution, we find that 87 are plausible, implying that the cut has a false positive rate of 2\%. We classify as plausible 17 of the 142 events which fail the cut, implying a false negative rate of 12\%. If a pure sample is required, we additionally recommend only considering events where both the source and lens have 2MASS detections. This leaves no false-positives whilst increasing the false-negative rate to 27\%. We note that for the two events we eliminated from \cite{Bramich2018}, both the source and lens have 2MASS detections.

\section{Conclusions}
We critically analyse the fidelity of predicted microlensing events extracted from GDR2. We find that a significant portion of the bright events ($G_{\text{source}}<18$) which are promising candidates for detection with \gaia are likely not genuine. This is demonstrated with a case study of G123-61, a high quality candidate for which two lenses are predicted to pass over the same source. Comparing with DSS and 2MASS observations we find that the source in both events is almost certainly not real and is generated in \gaia from misclassified observations of the two lens objects. We propose that sources brighter than $G=18$ should only be considered if they have a published 5D astrometric solution. We demonstrate that this cut increases the reliability of the sample of microlensing events and significantly changes the outlook for measuring precise lens masses with future \gaia data. We recommend this cut for all microlensing events searches with \gaia. On the positive side, our findings imply that at least half of the astrometric microlensing events during the \gaia extended mission are yet to be identified. In the online Table associated with this paper we list all of the events predicted using GDR2 by each of the studies we considered, with results of the visual inspection described above, 2MASS lens and source IDs where available and \gaia 5D astrometry flags.

\section*{Acknowledgements}

PM and AE would like to thank STFC for studentship funding. DB thanks Magdalen College for his fellowship and the Rudolf Peierls Centre for Theoretical Physics for providing office space and travel funds. The authors are grateful to Timo Prusti at the \gaia Helpdesk for useful discussions that enabled this work. We would like to thank the referee Ulrich Bastian for useful suggestions. This work presents results from the European Space Agency (ESA) space mission \gaia. \gaia data are being processed by the \gaia Data Processing and Analysis Consortium (DPAC). Funding for the DPAC is provided by national institutions, in particular the institutions participating in the \gaia MultiLateral Agreement (MLA). The \gaia mission website is \url{https://www.cosmos.esa.int/gaia}. The \gaia archive website is \url{https://archives.esac.esa.int/gaia}.

\section*{Data availability}

The data underlying this article are available in the article and in its online supplementary material.



\bibliographystyle{mnras}
\bibliography{references} 




\appendix


\bsp	
\label{lastpage}
\end{document}